\documentclass[british,english,a4paper,superscriptaddress,aps,twocolumn,pre]{revtex4-1}

\usepackage{amssymb,amsmath} 
\usepackage{graphicx}

\usepackage{appendix}
\usepackage[small,bf]{caption}

\usepackage{color}
\usepackage[usenames,dvipsnames,svgnames,table]{xcolor}
\begin{document}

\title{Strong geometric frustration in model glassformers}

\author{C. Patrick Royall}
\affiliation{HH Wills Physics Laboratory, Tyndall Avenue, Bristol, BS8 1TL, UK}
\affiliation{School of Chemistry, University of Bristol, Cantock Close, Bristol, BS8 1TS, UK}
\affiliation{Centre for Nanoscience and Quantum Information, Tyndall Avenue, Bristol, BS8 1FD, UK}

\author{Alex Malins}
\affiliation{School of Chemistry, University of Bristol, Cantock Close, Bristol, BS8 1TS, UK}
\affiliation{Bristol Centre for Complexity Science, University of Bristol, Bristol, BS8 1TS, UK}

\author{Andrew J. Dunleavy}
\affiliation{HH Wills Physics Laboratory, Tyndall Avenue, Bristol, BS8 1TL, UK}
\affiliation{School of Chemistry, University of Bristol, Cantock Close, Bristol, BS8 1TS, UK}
\affiliation{Bristol Centre for Complexity Science, University of Bristol, Bristol, BS8 1TS, UK}
\affiliation{Centre for Nanoscience and Quantum Information, Tyndall Avenue, Bristol, BS8 1FD, UK}

\author{Rhiannon Pinney}
\affiliation{HH Wills Physics Laboratory, Tyndall Avenue, Bristol, BS8 1TL, UK}
\affiliation{Bristol Centre for Complexity Science, University of Bristol, Bristol, BS8 1TS, UK}

\email{paddy.royall@bristol.ac.uk}

\date{\today}

\begin{abstract}
We consider three popular model glassformers, the Kob-Andersen and Wahnstr\"{o}m binary Lennard-Jones models and weakly polydisperse hard spheres. Although these systems exhibit a range of fragilities, all feature a rather similar behaviour in their local structure approaching dynamic arrest. In particular we use the dynamic topological cluster classification to extract a locally favoured structure which is particular to each system. These structures form percolating networks, however in all cases there is a strong decoupling between structural and dynamic lengthscales. We suggest that the lack of growth of the structural lengthscale may be related to strong geometric frustration.
\end{abstract}

\maketitle

\section{Introduction}

Among the challenges of the glass transition is how solidity emerges with little apparent change in structure ~\cite{berthier2011,royall2014physrep}. However, using computer simulation and
with the advent of particle-resolved studies in colloid experiments ~\cite{ivlev}, it has become possible to construct and use higher-order correlation functions ~\cite{tanemura1977,steinhardt1983,honeycutt1987,malins2013tcc}. These can directly identify local geometric motifs in supercooled liquids, long-since thought to suppress crystalisation in glassforming systems ~\cite{frank1952}. Other indirect approaches include the use of reverse Monte Carlo techniques to extract higher-order information from two-point correlation functions  ~\cite{mcgreevy2001} which is used in metallic glassformers  for example ~\cite{cheng2011}.

Such measurements have correlated the occurrence of geometric motifs and slow dynamics in a number of glassformers in both particle-resolved experiments on colloids ~\cite{vanblaaderen1995,konig2005,royall2008} and simulation ~\cite{jonsson1988,tomida1995,perera1999a,perera1999b,dzugutov2002}. Identification of these motifs 
has led to the tantalising prospect of finding a structural mechanism for dynamic arrest. It has been demonstrated that at sufficient supercooling, there should be a coincidence in structural and dynamic lengths, associated with regions undergoing relaxation for fragile glassformers ~\cite{montanari2006}. Thus recent years have seen a considerable effort devoted to identifying dynamic and structural lengthscales in a range of glassformers. The jury remains out concerning the coincidence of structural and dynamic lengthscales, 
with some investigations finding agreement between dynamic and structural lengthscales in experiment ~\cite{leocmach2012} and simulation 
~\cite{shintani2006,kawasaki2007,kawasaki2010jpcm,tanaka2010,sausset2010,mosayebi2010,mosayebi2012,xu2012}, while others find that the while the dynamic lengthscale increases quite strongly, structural correlation lengths grow weakly if at all ~\cite{karmakar2009,hocky2012,charbonneau2012,kob2011non,dunleavy2012,charbonneau2013,charbonneau2013pre,malins2013jcp,malins2013fara}.
Other interpretations include decomposing the system into geometric motifs and considering the motif system. One such effective system exhibits no glass transition at finite temperature ~\cite{eckmann2008}.

\begin{figure}[!htb]
\begin{center}
\includegraphics[width=45mm]{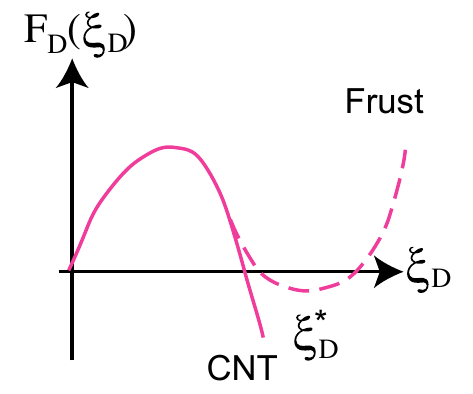} 
\caption{
\textbf{Schematic of geometric frustration limiting the growth of domains of locally favoured structures.} Solid line is conventional CNT with the first two terms in Eq. \ref{eqCNTFrustration} which would occur in the non-frustrated case. Dashed line denotes the effect of the third term which incorporates frustration, leading to a preferred lengthscale for the LFD domain 
 $\xi_D^*$.\label{figFrustration}}
\vspace*{-12pt}\end{center}
\end{figure}

Here we consider the approach of geometric frustration ~\cite{tarjus2005}. Geometric frustration posits that upon cooling, a liquid will exhibit an increasing number of locally favoured structures (LFS), which minimise the local free energy. In some unfrustrated curved space, these LFS tessellate, and there is a phase transition to an LFS-phase. In Euclidean space, frustration limits the growth of the LFS domains. As detailed in section \ref{sectionFrustration}, the free energy associated with the growth of these LFS domains may be related to an addition term to classical nucleation theory (CNT), as illustrated schematically in Fig. \ref{figFrustration}.

Now in 2d monodisperse hard discs, the locally favoured structure (hexagonal order) is commensurate with the crystal. The transition is weakly first order to a hexatic phase which exhibits a  continuous transition with the 2d crystal ~\cite{bernard2011,engel2013}. Thus in 2d one must curve space to introduce geometric frustration. This has been carried out by Sausset \emph{et al.} \cite{sausset2008}, curving in hyperbolic space, where the degree of curvature can be continuously varied. Weakly curved systems have a strong tendency to hexagonal ordering, which was controllably frustrated by the curvature. However, the upper bound on all correlation lengths was dictated by the curvature in this system. In other words, frustration is encoded into the system through the curved space, suppressing any divergent structural lengthscales. However structural lengthscales were observed to grow up to the limit set by the curved space ~\cite{sausset2010,sausset2010pre}. In 3d, 600 perfect (strain-free) tetrahedra formed from 120 particles can be embedded on the surface of a four-dimensional sphere ~\cite{coexeter,coexeterpolytope}. Each particle in this 4d Platonic solid or ``polytope'' is at the centre of a 12-particle icosahedrally coordinated shell, and indeed simulations indicate a continuous transition in this system ~\cite{nelson1983,nelson,straley1984}. However,  a 120 particle system is clearly inappropriate to any investigation of increasing lengthscales.

Here we focus on geometric frustration in 3d Euclidean space. We carry out simulations on a number of well-known glassformers of varying degrees of fragility: polydisperse hard spheres, and the Kob-Andersen~\cite{kob1995a} and Wahnstr\"{o}m binary Lennard-Jones mixtures ~\cite{wahnstrom1991}. In each system we identify a system-specific locally favoured structure ~\cite{malins2013tcc}, which becomes more prevalent as the glass transition is approached. We measure the dynamic correlation length $\xi_4$ and identify a structural correlation length $\xi_{\mathrm{LFS}}$ ~\cite{malins2013jcp,malins2013fara}. We show that the dynamic correlation length grows much more than the structural correlation length associated with the LFS in each system. The LFS do not tile 3d space, but instead form system spanning networks. We conclude that the growth of LFS in all these cases is strongly frustrated. Given the system specific nature of the LFS, we speculate that an LFS-phase might not in principle require curved space and we suggest that geometric frustration might be considered not as a function of curvature but as composition.

This paper is organised as follows. In section \ref{sectionFrustration} we briefly consider some pertinent aspects of geometric frustration theory, followed by a description of our simulations in section \ref{sectionMethods}. The results consist of a connection between the fragility of the systems studied placed in the context of some molecular glassformers in section \ref{sectionFragility}. In section \ref{sectionIdentifying} we detail how the locally favoured structures are identified and discuss the increase in LFS in section \ref{sectionFraction}. In section \ref{sectionStaticDynamic} we show structural and dynamic correlation lengths, before discussing our findings in section \ref{sectionDiscussion} and concluding in section \ref{sectionConclusions}.

\section{Geometric Frustration}
\label{sectionFrustration}

For a review of geometric frustration, the reader is directed to Tarjus \emph{et al.} \cite{tarjus2005}. The effects of frustration upon a growing domain of locally favoured structures may be considered as defects, which typically interact in a Coulombic fashion. Under the assumpton that frustration is \emph{weak}, this argument leads to scaling relations for the growth of domains of LFS, whose (linear) size we denote as $\xi_{D}$. Weak frustration requires that its effects only become apparent on lengthscales larger than the constituent particle size such that $\sigma << \xi_{D}$. Geometric frustration imagines an avoided critical point, at $T_x^c$, which corresponds to the phase transition to an LFS state in the unfrustrated system. At temperatures below this point, growth of domains of the LFS in the frustrated system may follow a classical nucleation theory (CNT) like behaviour, with an additional term to account for the frustration. In $d=3$ the free energy of formation of a domain size $\xi_{D}$ of locally favoured structures thus reads

\begin{equation}
F_{D}(\xi_{D},T)= \gamma(T)\xi_{D}^\theta-\delta \mu (T) \xi_{D}^3 + s_{\mathrm{frust}}(T)\xi_{D}^5
\label{eqCNTFrustration}
\end{equation}

\noindent where the first two terms express the tendency of growing locally preferred order and they represent, respectively, the energy cost of having an interface between two
phases and a bulk free-energy gain inside the domain. Equation \ref{eqCNTFrustration} is shown schematically in Fig. \ref{figFrustration}. The value of $\theta$ may be related to Adam-Gibbs theory ~\cite{adam1965} or Random First Order Transition theory (RFOT) ~\cite{lubchenko2007}. Without the third term long-range order sets in at $T =T_x^c$, in the unfrustrated system. Geometric frustration is encoded in the third term which represents the strain free energy resulting from the frustration. This last term is responsible for the fact that the transition is avoided and vanishes in the limit of non-zero frustration ~\cite{tarjus2005}. While actually evaluating the coefficients in Eq. \ref{eqCNTFrustration} is a very challenging undertaking, one can at least make the following qualitative observation. In the case of weak frustration, one expects extended domains of LFS. However, in the case of strong frustration, one imagines rather smaller domains of LFS, as the third term in Eq. \ref{eqCNTFrustration} will tend to dominate.

\section{Simulation Details}
\label{sectionMethods}

Our hard sphere simulations use the DynamO package ~\cite{bannerman2011}. This performs event-driven MD simulations, which we equilibrate for 300$\tau_\alpha$, in the NVT ensemble, before sampling in the NVE ensemble. We use two system sizes of $N=1372$ and  $N=10976$ particles, in a five-component equimolar mixture whose diameters are $[0.888\sigma, 0.95733\sigma, \sigma, 1.04267\sigma, 1.112\sigma]$, which corresponds to a polydispersity of 8\%. Here $\sigma$ is a diameter which we take as the unit of length. We have never observed crystallisation in this system. Given the moderate polydispersity, we do not distinguish between the different species. We use smaller systems of $N=1372$ to determine the structural relaxation time and the fraction of particles in locally favoured structures. Static and dynamic lengths are calculated for larger systems of $N=10976$. Further details may be found in ref. ~\cite{dunleavySub}.

We also consider the Wahnstr\"{o}m \cite{wahnstrom1991} and Kob-Andersen \cite{kob1995a} models in which the two species of Lennard-Jones particles interact with a pair-wise potential,  

\begin{equation}
u_{\mathrm{LJ}}(r) = 4 \epsilon_{\alpha \beta}\left[\left(\frac{\sigma_{\alpha\beta}}{r_{ij}}\right)^{12}-\left(\frac{\sigma_{\alpha\beta}}{r_{ij}}\right)^6\right]
\label{eqLJ}
\end{equation}

\noindent where $\alpha$ 
and $\beta$ denote the atom types $A$ and $B$, and $r_{ij}$ is the separation. In the equimolar Wahnstr\"{o}m mixture, the energy, length and mass values are $\varepsilon_{AA}=\epsilon_{AB}=\varepsilon_{BB}$, $\sigma_{BB}/\sigma_{AA}=5/6$, $\sigma_{AB}/\sigma_{AA}=11/12$ and $m_A=2m_B$ respectively. The simulations are carried out at a number density of $\rho=1.296$.
The Kob-Andersen binary mixture is composed of 80\% large (A) and 20\% small (B) particles possessing the same mass $m$~\cite{kob1995a}. The nonadditive Lennard-Jones interactions between each species, and the cross interaction, are given by $\sigma_\mathrm{AA}=\sigma$, $\sigma_\mathrm{AB}=0.8\sigma$, $\sigma_\mathrm{BB}=0.88\sigma$, $\epsilon_\mathrm{AA}=\epsilon$, $\epsilon_\mathrm{AB}=1.5\epsilon$, and $\epsilon_\mathrm{BB}=0.5\epsilon$ and is simulated at $\rho=1.2$.  For both Lennard-Jones mixtures, we simulate a system of $N=10976$ particles for an equilibation period of $300\tau_\alpha^A$ in the NVT ensemble and sample for a further $300\tau_\alpha^A$ in the NVE ensemble. The results are quoted in reduced units with respect to the A particles, \emph{i.e.} we measure length in units of $\sigma$, energy in units of $\epsilon$, time in units of $\sqrt{m\sigma^2/\epsilon}$, and set Boltzmann's constant $k_\mathrm{B}$ to unity.  Further details of the simulation of the Wahnstr\"{o}m and Kob Andersen models may be found in \cite{malins2013jcp,malins2013fara} respectively.

The $\alpha$-relaxation time $\tau_{\alpha}^{A}$ for each state point is defined by fitting the Kohlrausch-Williams-Watts stretched exponential to the alpha-regime of the intermediate scattering function (ISF) of the $A$-type particles in the case of the Lennard-Jones mixtures and of all particles in the case of the hard spheres.

\section{Results}
\label{sectionResults}

\subsection{Fragility}
\label{sectionFragility}

\begin{figure}[!htb]
\begin{center}
\includegraphics[width=80mm]{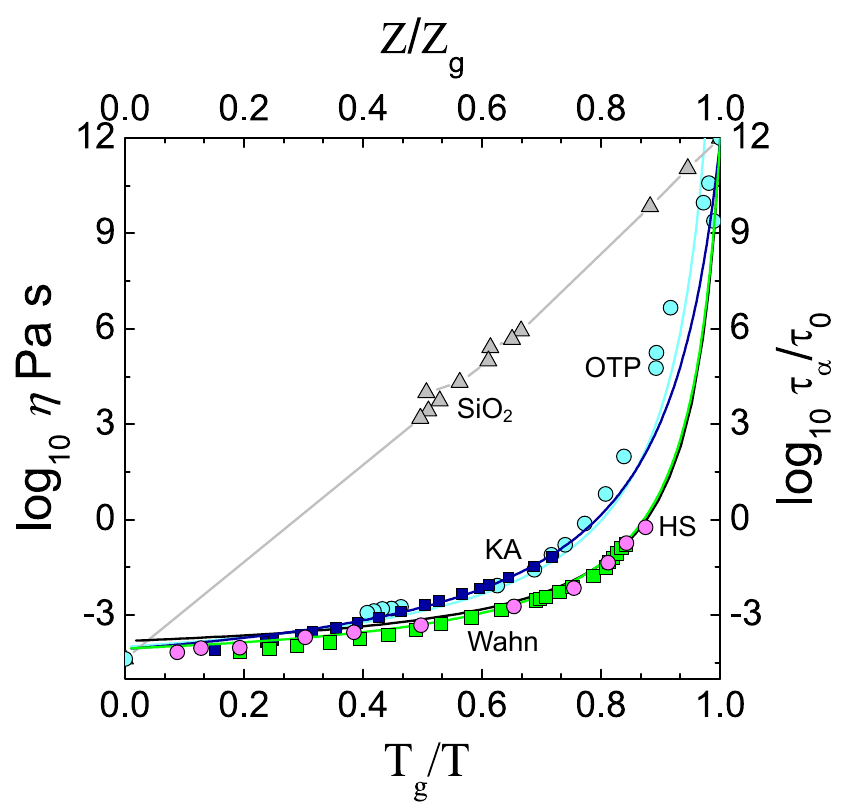} 
\caption{\textbf{The Angell plot.} Data are fitted to VFT  (Eq. \ref{eqVFT})
and Eq.  \ref{eqVFTHS}) in the case of hard spheres. HS denotes hard spheres where the control parameter is reduced pressure $Z$. The ``experimental glass transition'' for hard spheres $\phi_g$ and the Lennard-Jones models are defined in the text. Data for SiO$_2$ and orthorterphenyl (OTP) are quoted from Angell ~\cite{angell1995} and 
Berthier and Witten ~\cite{berthier2009witten}. For the systems studied here, $\tau_0$ is scaled to enable data collapse at $T_g/T = \phi/\phi_g \rightarrow 0$.}
\label{figAngell}
\vspace*{-12pt}\end{center} 
\end{figure}

We fit the structural relaxation time to the Vogel-Fulcher-Tamman (VFT) equation

\begin{equation}
\tau_{\alpha}=\tau_{0} \exp \left[ \frac{A}{(T-T_{0})} \right].
\label{eqVFT}
\end{equation}

\noindent Here $\tau_0$ is a reference relaxation time, the parameter $A$ is related to the fragility $D=A/T_0$ and $T_0$ is temperature of the ``ideal'' glass transiton, at which $\tau_{\alpha}$ diverges. Of course experimental systems cannot be equilibrated near $T_0$, so the experimental glass transition $T_g$ is defined where $\tau_{\alpha}$ exceeds $100$ s in molecular liquids. In the Angell plot in Fig. \ref{figAngell} silica and ortho-terphenyl are fitted with Eq. \ref{eqVFT}, along with our data for the Wahnstr\"{o}m and Kob-Andersen mixtures. For our data, we obtain an estimate of $T_g$ as the temperature at which $\tau_\alpha /\tau_0 \approx 10^{15}$ with the VFT fit. We apply the VFT fit only for $T<1$, which denotes the onset temperature for the activated dynamics in which VFT is appropriate. Higher temperatures exhibit an Arrhenius-like behaviour ~\cite{malins2013jcp,malins2013fara}. The fragility is then $D=A/T_0$. Details of the fitted, and literature ~\cite{angell1995,richert1998,berthier2009witten} values are given in Table \ref{tableFragility}. 

In the case of hard spheres, temperature plays no role and packing fraction $\phi$ is typically used as a control parameter, not least as it may be measured in experiments, albeit with limited precision ~\cite{poon2012,royall2013myth}. However convincing arguments have been made by Berthier and Witten ~\cite{berthier2009witten} and by Xu \emph{et al.} ~\cite{xu2009} that the reduced pressure $Z=p/(\rho k_BT)$ where $p$ is the pressure is in fact more analogous to $1/T$ in molecular systems. One result of this observation is that $Z$ diverges at (random) close packing, forming an analogy with the zero-temperature limit. In addition, the VFT form may be generalised with an exponent $\delta$ in the denominator ~\cite{berthier2009witten,brambilla2009}. 

\begin{equation}
\tau_{\alpha}=\tau_{0} \exp\left[ \frac{A}{(Z_{0}-Z)^\delta} \right].
\label{eqVFTHS}
\end{equation}

We have investigated fitting with $Z$ and with $\phi$, and in the case that $\delta=1$ have found little difference if we make the significant assumption that the Carnahan-Starling equation of state holds for $Z(\phi)$. Like Berthier and Witten ~\cite{berthier2009witten}, we find good agreement with Carnahan-Starling at all state points accessible to simulation. We have also investigated setting $\delta=2.2$, where we find $Z_0 = 45.39$, a value at which the Carnahan Starling relation corresponds to a value of $\phi_0\approx0.66$ which is slightly greater than random close packing. This difference with previous work ~\cite{berthier2009witten,brambilla2009}, may reflect our choice of hard sphere system. At 8\%, ours is weakly polydisperse. In any case, we have found a better fit over a larger range of $Z$ when $\delta=1$, and quote those values in Table  \ref{tableFragility}. The value of $Z_0$ we obtain corresponds, via the Carnahan-Starling relation to $\phi_0=0.603$. We  also define a  $Z_g$ using the VFT equation for hard spheres (Eq. \ref{eqVFTHS}) in an analogous way to that which we used to determine $T_g$ for the Lennard-Jones mixtures. 

As the values in Table \ref{tableFragility} and Fig. \ref{figAngell} show, the Kob-Andersen, hard sphere and Wahnstr\"{o}m systems exhibit progressively higher fragilities. In Fig. \ref{figAngell}, the Kob-Andersen system sits close to ortho-terphenyl in the range of supercoolings accessible to our simuations. The fragility of the former we find to be 3.62, while the latter is quoted to be around 10 \cite{berthier2009witten,richert1998}. For our VFT fit to the Kob-Andersen mixture we took a literature value ~\cite{speck2012} of $T_0=0.325$, however a free fit of our data leads to a fragility $D\approx7.06$, close to the orthoterphenyl (OTP) value. Since OTP is at the fragile end of molecular glassformers ~\cite{angell1995}, Fig. ~\ref{figAngell} suggests that the models considered here are fragile when compared to molecular systems.

We emphasise that fitting VFT is not the only approach by any means \cite{hecksler2008}. In particular it is possible to consider an energy of activation $E(T)$ which is given by an Arrhenius form $\tau_\alpha \propto \exp[\beta E(T)]$ ~\cite{tarjus2000}. This approach has been carried out for both Lennard-Jones mixtures considered here (in the isobaric-isothermal ensemble) ~\cite{coslovich2007ii} where KA was again found to be less fragile than the Wahnstr\"{o}m model. Both were found to be less fragile than molecular glassformers, which could reflect the limited dynamically accessible range. Here on the other hand we have chosen to extrapolate our simulation data to much larger dynamic ranges.

\begin{table*}
  \begin{tabular}{ | c | c | c | c | c | c | } \hline
 system & $T_0$, $Z_0$ & $T_g$, $\phi_g$  & $D$ & LFS & reference \\ \hline
SiO$_2$ &  * & 820-900K & $\sim$60 & & \cite{berthier2009witten}  \\
OTP & 202K & 246K & $\sim$10 & & \cite{richert1998} \\
KA & 0.325 & 0.357$\pm$0.005 & 3.62$\pm$0.08 & 11A & \cite{malins2013fara}, this work \\
Wahn & 0.464$\pm$0.007 & 0.488$\pm$0.005 & 1.59$\pm$0.13 & 13A & \cite{malins2013jcp}, this work \\
HS & 28.0$\pm$1.2 & 26.8$\pm$1.0 & 1.711$\pm$0.54 & 10B & \cite{taffs2013}, this work \\  \hline
\end{tabular}
  \caption{Transition temperatures, fragilities and locally favoured structures for systems in Fig. \ref{figAngell}. KA denotes Kob-Andersen mixture, Wahn Wahnstr\"{o}m mixture and HS hard spheres. Note that, as a strong liquid, the fragility of silica is poorly defined ~\cite{angell1995}.}\label{tableFragility}
\end{table*}

\subsection{Identifying the locally favoured structure}
\label{sectionIdentifying}

\begin{figure*}[!htb]
\begin{center}
 \includegraphics[width=120mm]{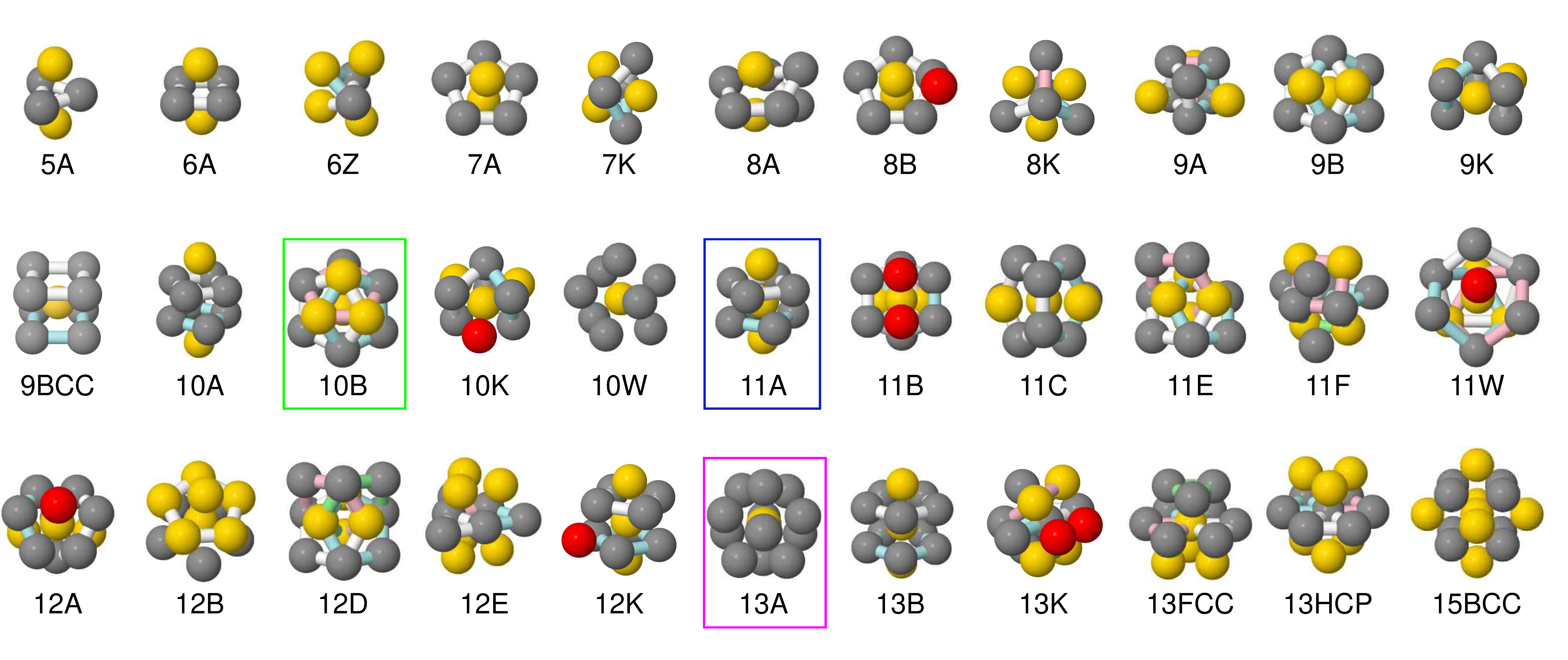} 
\caption{\textbf{The structures detected by the topological cluster classification ~\cite{malins2013tcc}}. Letters correspond to different models, numbers to the number of atoms in the cluster. K is the Kob-Andersen model ~\cite{malins2013fara}, W is the Wahnstr\"{o}m model ~\cite{malins2013jcp}. Outlined are the locally favoured structures identified for the Kob-Andersen model (11A), Wahnstr\"{o}m model (13A) and hard spheres (10B). Other letters correspond to the variable-ranged Morse potential, letters at the start of the alphabet to long-ranged interactions, later letters to short-ranged interactions, following Doye \emph{et al.} \cite{doye1995}. Also shown are common crystal structures.}
\label{figTCC} 
\vspace*{-12pt}\end{center}
\end{figure*}

In order to identify locally favoured structures relevant to the slow dynamics, we employ the dynamic topological cluster classification algorithm ~\cite{malins2013jcp,malins2013fara}. This measures the lifetimes of different clusters identified by the topological cluster classification (TCC) ~\cite{malins2013tcc}. The TCC identifies a number of local structures as shown in Fig. \ref{figTCC}, including those which are the minimum energy clusters for $m=5$ to $13$ Kob-Andersen ~\cite{malins2013fara} and Wahnstr\"{o}m ~\cite{malins2013jcp} particles in isolation. In the case of the hard spheres, minimum energy clusters are not defined. However we have shown that the Morse potential, when truncated at its minimum in a similar fashion to the Weeks-Chandler-Andersen treatment for the Lennard-Jones model ~\cite{weeks1971} provides an extremely good approximation to hard spheres ~\cite{taffs2010jcp}. Clusters corresponding to the (full) Morse potential have been identified by Doye \emph{et al.} ~\cite{doye1995} are included in the TCC. The first stage of the TCC algorithm is to identify the bonds between neighbouring particles. The bonds are detected using a modified Voronoi method with a maximum bond length cut-off of $r_\mathrm{c}=2.0$ ~\cite{malins2013tcc}. 

In the case of the Kob-Andsersen mixture, a parameter which controls identification of four- as opposed to three-membered rings $f_\mathrm{c}$ is set to unity thus yielding the direct neighbours of the standard Voronoi method. Under these conditions, 11A bicapped square antiprism clusters are identified ~\cite{malins2013fara,malins2013tcc}, which have previously been found to be important in the Kob-Andersen mixture ~\cite{coslovich2007}. For the Wahnstr\"{o}m mixture and the hard spheres, the four-membered ring parameter $f_\mathrm{c}=0.82$ which has been found to provide better discrimination of long-lived icosahedra ~\cite{malins2013jcp}.

In the dynamic TCC, a lifetime $\tau_{\ell}$ is assigned to each ``instance'' of a cluster, where an instance is defined by the unique indices of the particles within the cluster and the type of TCC cluster. Each instance of a cluster occurs between two frames in the trajectory and the lifetime is the time difference between these frames. We require that no subset of the particles becomes un-bonded from the others during the lifetime of the instance, i.e. we require that the same particles comprise the cluster though out its lifetime. However due to bond breaking from thermal fluctuations, sometimes the cluster bond topology can change. Such periods are constrained to be less than $\tau_\alpha$ in length. The longest lived clusters detected in this way we interpret as locally favoured structures ~\cite{malins2013jcp,malins2013fara}.

The measurement of lifetimes for all the instances of clusters is intensive in terms of the quantity of memory required to store the instances, and the number of searches through the memory required by the algorithm each time an instance of a cluster is found to see if it existed earlier in the trajectory. Therefore we do not measure lifetimes for the clusters where $N_\mathrm{c}/N>0.8$, since the vast majority of particles are found within such clusters and it is not immediately clear how dynamic heterogeneities could be related to structures that are pervasive throughout the whole liquid. Here $N_\mathrm{c}/N$ is the fraction of particles in the system which are included in a given cluster. For example five particles are included in 5A and 13 in the 13A icosahedron. We do not distinguish between particle types, and treat all parts of a cluster on an equal footing. Some particles can be included in more than one cluster because clusters (of the same type can overlap), but these only count once towards $N_\mathrm{c}/N$.

\begin{figure*}
\begin{center}
\includegraphics[width=7cm,keepaspectratio]{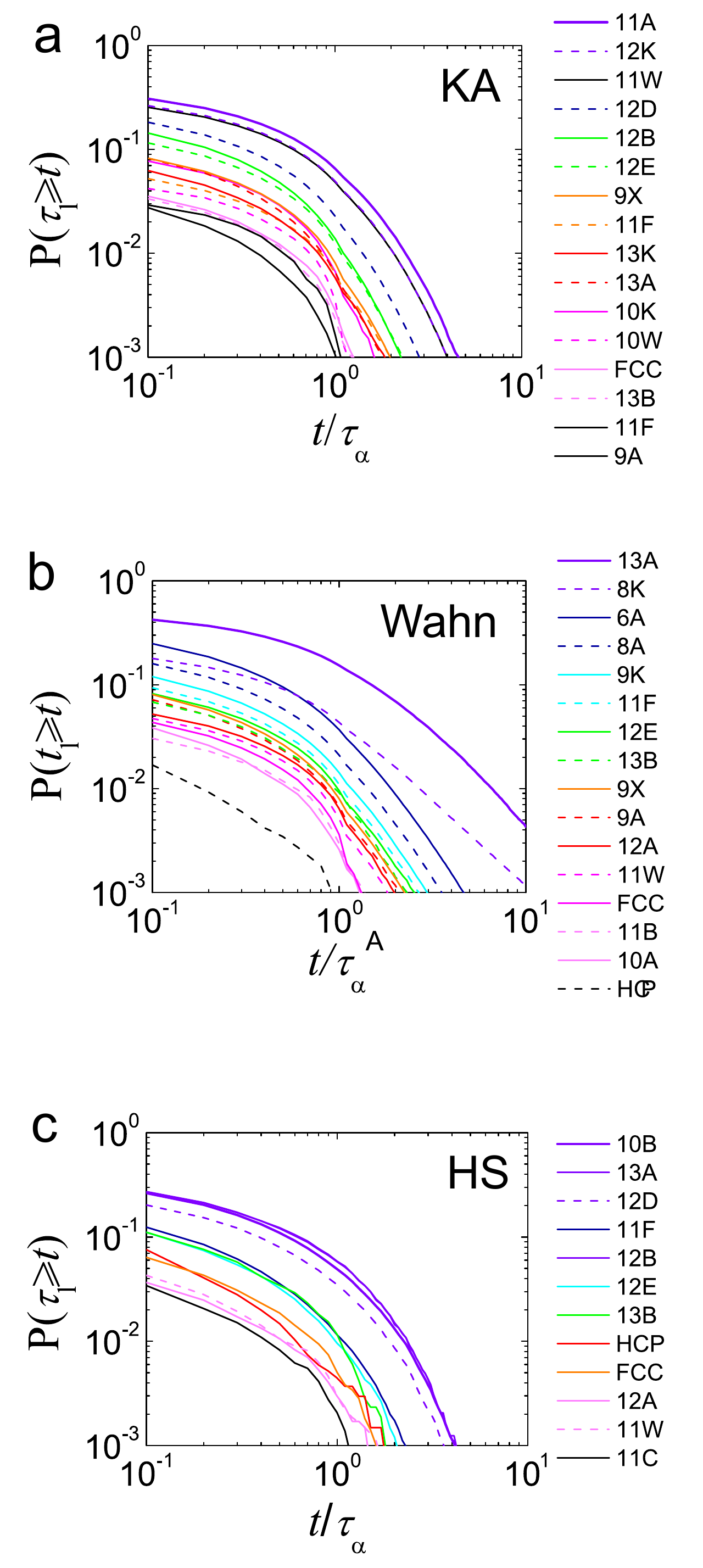}
 \caption{Lifetime autocorrelation functions for the TCC clusters $P(\tau_{\ell} \ge t)$.
(a) Kob-Andersen mixture $T=0.498$, (b) Wahnstr\"{o}m mixture $T=0.606$ and (c) hard spheres $Z=25.1$ ($\phi=0.57$).
Particle colours show how the cluster is detected by the TCC ~\cite{malins2013tcc}.}
\label{figDTCC}
\vspace*{-12pt}\end{center}
\end{figure*}

We plot the results of the dynamic TCC in Fig. \ref{figDTCC} for a low temperature state point for the Lennard-Jones mixtures and for $\phi=0.57$ in the case of hard spheres. This clearly shows the most persistent or the longest lived of the different types of clusters in each system. All three systems exhibit rather similar behaviour, namely that the long-time tail of the  autocorrelation function indicates that some of these clusters preserve their local structure on timescales far longer than $\tau_{\alpha}$, and these we identify as the LFS. Thus for the Kob-Andersen mixture, we identify 11A bicapped square antiprisms ~\cite{malins2013fara}, for the Wahnstr\"{o}m model 13A icosahedra ~\cite{malins2013jcp} and for hard spheres 10B. In the hard sphere case, other clusters are also long-lived : 13A, 12B and 12D. However these are only found in small quantities ($\lesssim 2$\%), unlike 10B which can account for up to around 40\% of the particles in the system. Moreover, a 10B cluster is a 13A cluster missing three particles from the shell, thus all 13A also correspond to 10B by construction. Related observations have been made concerning the Kob-Andersen ~\cite{malins2013fara} and Wahnstr\"{o}m mixtures ~\cite{malins2013jcp}. 

The fast initial drops of $P(\tau_\ell \ge t)$ reflect the existence of large numbers of clusters with lifetimes $\tau_\ell \ll \tau_\alpha$. The lifetimes of these clusters are comparable to the timescale for $\beta$-relaxation where the particles fluctuate within their cage of neighbours. It could be argued that these clusters arise spuriously due to the microscopic fluctuations within the cage, and that the short-lived clusters are not representative of the actual liquid structure. However almost no LFS are found at higher temperatures (or lower volume fraction in the case of hard spheres), cf. Fig. \ref{figNLFSN}, where microscopic fluctuations in the beta-regime also occur. We have not yet found a way to distinguish between the short and long-lived LFS structurally, so we conclude that the measured distribution of LFS lifetimes, which includes short-lived clusters, is representative of the true lifetime distribution.

\subsection{Fraction of particles participating within LFS}
\label{sectionFraction}

\begin{figure*}
\begin{center}
\includegraphics[width=12cm,keepaspectratio]{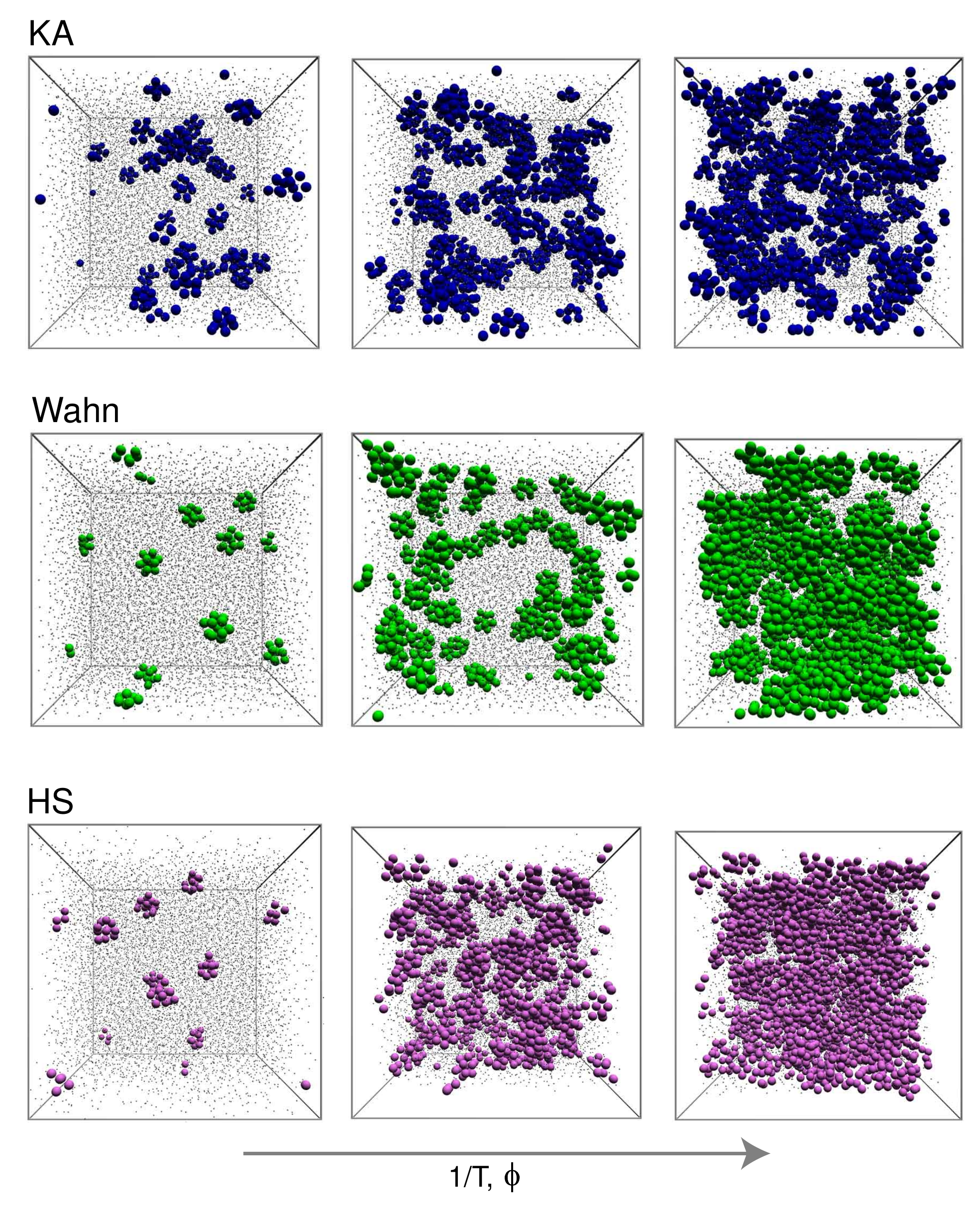}
\caption{\textbf{Snapshots of locally favoured structures in the three systems considered.} In all cases particles identified in LFS are drawn to 80\% actual size and coloured, other particles are rendered at 10\% actual size and are grey. 
Top row : Kob-Andersen mixture for $T=1$, $0.6$ and $0.5$ from left to right.
Middle row :  Wahnstr\"{o}m mixture for $T=1.0$, $0.625$, $0.606$ from left to right. 
Bottom row : Hard spheres for $Z=6.92$, $Z=13$, $Z=18.5$ ($\phi=0.4$, $0.5$, $0.55$) from left to right.}
\label{figPretty}
\vspace*{-12pt}\end{center}
\end{figure*}

Having identified the locally favoured structure for each system, we consider how the particles in the supercooled liquids are structured using the topological cluster classification algorithm ~\cite{malins2013tcc}. We begin with the snapshots in Fig. \ref{figPretty}. It is immediately clear that the spatial distribution of the LFS is similar in all three systems. In all cases, at weak supercooling isolated LFS appear, becoming progressively more popular upon deeper supercooling. At the deeper quenches, the LFS percolate, but the ``arms'' of the percolating network are around three-four particles thick. One caveat to this statement is that at high temperature, the Kob-Andersen mixture exhibits more LFS than a comparable state point in the Wanstr\"{o}m mixture. Furthermore, the geometry of the LFS domains is clearly much more complex than spherical nuclei assumed in classical nucleation theory (Fig. ~\ref{figFrustration}). Indeed one might imagine that deconstructing such a complex structure to a single linear length may warrant consideration.

\begin{figure*}
\begin{center}
\includegraphics[width=12cm,keepaspectratio]{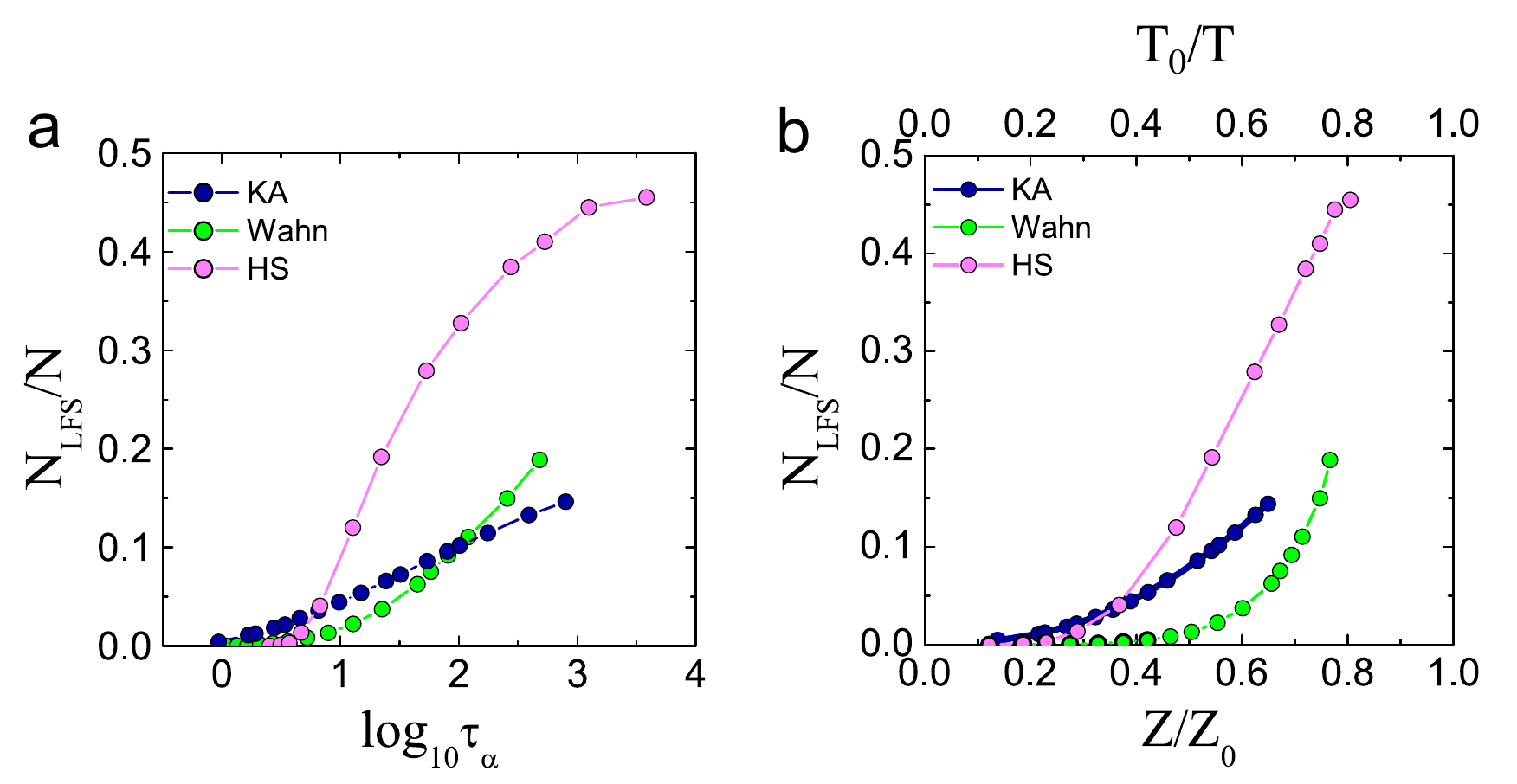}
 \caption{\textbf{Fraction of particles in locally favoured structures $N_{\mathrm{LFS}}/N$.}
(a) $N_{\mathrm{LFS}}/N$ as a function of $\tau_\alpha/\tau_0$.
(b) $N_{\mathrm{LFS}}/N$ as a function of $T_0/T$ and $Z/Z_0$ for the Lennard-Jones and hard sphere
systems respectively.
In (a) $\tau_\alpha/\tau_0$ are offset for clarity.}
\label{figNLFSN}  
\vspace*{-12pt}\end{center}
\end{figure*}

In Fig. \ref{figNLFSN} we plot the fraction of particles detected within LFS for each system $N_{\mathrm{LFS}}/N$. We consider the scaled structural relaxation time $\tau_\alpha/\tau_0$ (see Fig. \ref{figAngell}) in Fig. \ref{figNLFSN}(a). In Fig. \ref{figNLFSN}(b) we show the population of LFS as a function of the degree of supercooling, $T_0/T$ and $Z/Z_0$ for the Lennard-Jones and hard sphere systems respectively. We find that the hard spheres show a dramatic increase in LFS, which appears to begin to level off for high values of $\tau_\alpha$. Note that by construction, $N_{\mathrm{LFS}}/N \leq 1$. This levelling off has recently been observed in biased simulations of the Kob-Andersen system, which exhibits a first-order transition in trajectory space to a dynamically arrested LFS-rich phase ~\cite{speck2012}. Such a levelling off may also be related to a fragile-to-strong transition observed in certain metallic glassformers ~\cite{wei2013}. Fragility has been correlated with a significant degree of structural change ~\cite{ito1999}. If the population of LFS somehow saturates (in any case, $N_\mathrm{LFS}$ cannot exceed unity), then the structure may change little upon deeper supercooling and it is possible that a crossover to strong behaviour may be observed.

Our simulations of the two Lennard-Jones systems do not reach such deep supercooling, so we have not yet determined whether they exhibit the same behaviour. However the increase of LFS in the case of the Kob-Andersen mixture is rather slower than the hard spheres, and the Wahnstr\"{o}m mixture is intermediate between the two. This correlates with the fragilities of these three systems, Fig. ~\ref{figAngell}. This connection between structure and fragility has been previously noted in the case of the two Lennard-Jones mixtures  ~\cite{coslovich2007}. 

Plotting as a function of supercooling, in Fig.  \ref{figNLFSN}(b), we find that the Kob-Andersen mixture shows a slow increase in LFS population which begins at quite weak supercooling, while hard spheres (recalling that here the control parameter is $Z$) show a much more rapid growth which begins at much deeper supercooling. 
The growth in LFS population in the Wahnstr\"{o}m mixture begins at deeper supercooling than the KA, however the population growth is then quite rapid. 

\subsection{Static and dynamic lengthscales}
\label{sectionStaticDynamic}

\begin{figure*}
\begin{center}
\includegraphics[width=14cm,keepaspectratio]{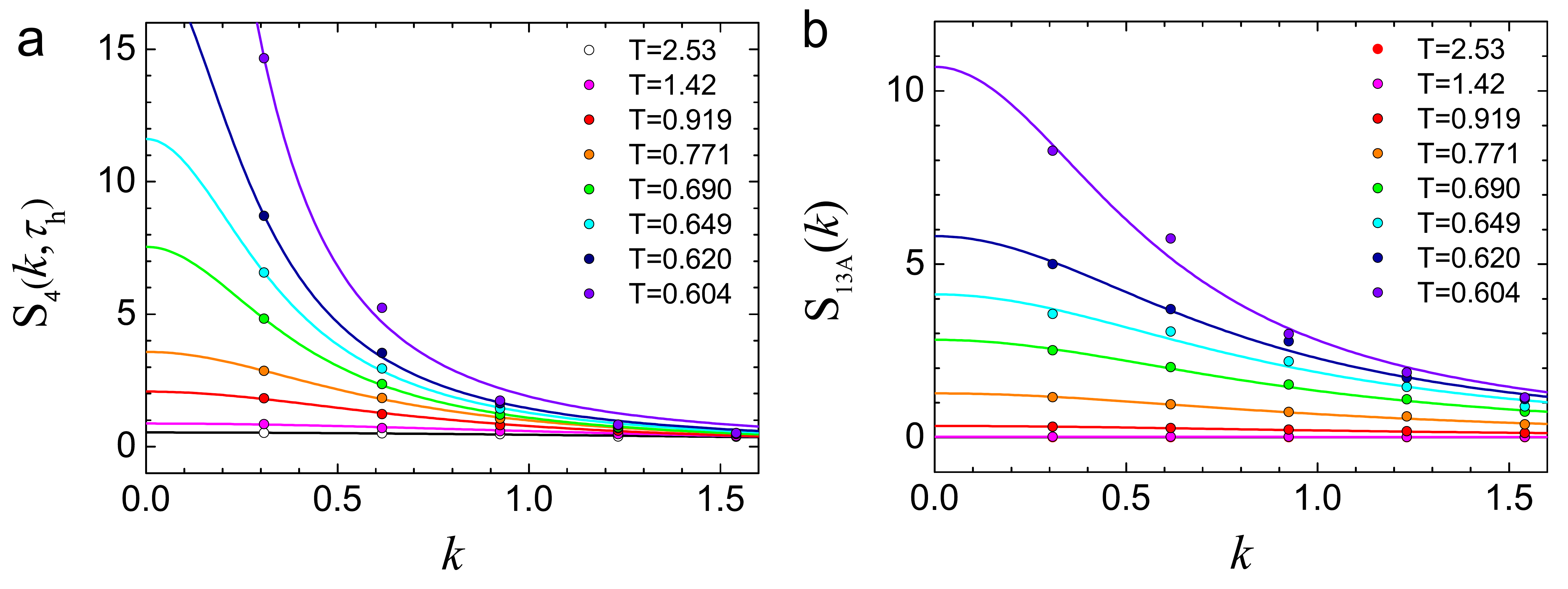}
 \caption{\textbf{Example fits to extract static and dynamic correlation lengths.}
(a) Fitting Eq. \ref{eqOZ} to obtain $\xi_4$. 
(b) Fitting Eq. \ref{eqSLFS} to obtain $\xi_\mathrm{LFS}$. Both (a) and (b) show data for the Wahnstr\"{o}m model.}
\label{figS}
\vspace*{-12pt}\end{center}
\end{figure*}

\textit{Dynamic correlation length --- }
We now turn to the topic with which we opened this article, the coincidence or otherwise of static and dynamic lengthscales in these systems.  In order to do this, we calculate both, beginning with the dynamic correlation length $\xi_{4}$, following La\v{c}evi\'{c} \textit{et al.}~\cite{lacevic2003}. We provide a more extensive description of our procedure  elsewhere~\cite{malins2013jcp}. Briefly, the dynamic correlation length $\xi_4$ is obtained as follows. A (four-point) dynamic susceptibility is calculated as
\begin{equation}
\chi_4(t)=\frac{V}{N^2 k_B T}[\langle Q(t)^2\rangle -\langle Q(t)\rangle^2],
\label{eqChi4}
\end{equation}
\noindent where 
\begin{equation}
Q(t)=\frac{1}{N} \sum_{j=1}^N \sum_{l=1}^N w(|\textbf{r}_j(t+t_0)-\textbf{r}_l(t_0)|).
\label{eqQ}
\end{equation}
\noindent The overlap function $w(|\textbf{r}_j(t+t_0)-\textbf{r}_l(t_0)|)$ is defined to be unity if $|\textbf{r}_j(t+t_0)-\textbf{r}_l(t_0)|\le a$, 0 otherwise, where $a=0.3$. The dynamic susceptibility $\chi_4(t)$ exhibits a peak at $t=\tau_h$, which corresponds to the timescale of maximal correlation in the dynamics of the particles.
We then construct the four-point dynamic structure factor $S_4(\textbf{k},t)$:
\begin{eqnarray*}
S_4(\textbf{k},t)&=&\frac{1}{N\rho} \langle \sum_{jl} \exp[-i \textbf{k} \cdot \textbf{r}_l(t_0)]w(|\textbf{r}_j(t+t_0)-\textbf{r}_l(t_0)|)\\
&\times& \sum_{mn} \exp[i \textbf{k} \cdot \textbf{r}_n(t_0)]w(|\textbf{r}_m(t+t_0)-\textbf{r}_n(t_0)|) \rangle,
\end{eqnarray*}
\noindent where $j$, $l$, $m$, $n$ are particle indices and $\mathbf{k}$ is the wavevector. For time $\tau_h$, the angularly averaged version is $S_4(k,\tau_h)$. The dynamic correlation length $\xi_4$ is then calculated by fitting the Ornstein-Zernike (OZ) function to $S_4(k,\tau_h)$, as if the system were exhibiting critical-like spatio-temporal density fluctuations, 
\begin{equation}
S_4(k,\tau_h)=\frac{S_4[0,\tau_h)}{1+(k\xi_4(\tau_h)]^2},
\label{eqOZ}
\end{equation}
\noindent to $S_4(k,\tau_h)$ for $k<2$~\cite{lacevic2003}.

\begin{figure*}
\begin{center}
\includegraphics[width=12cm,keepaspectratio]{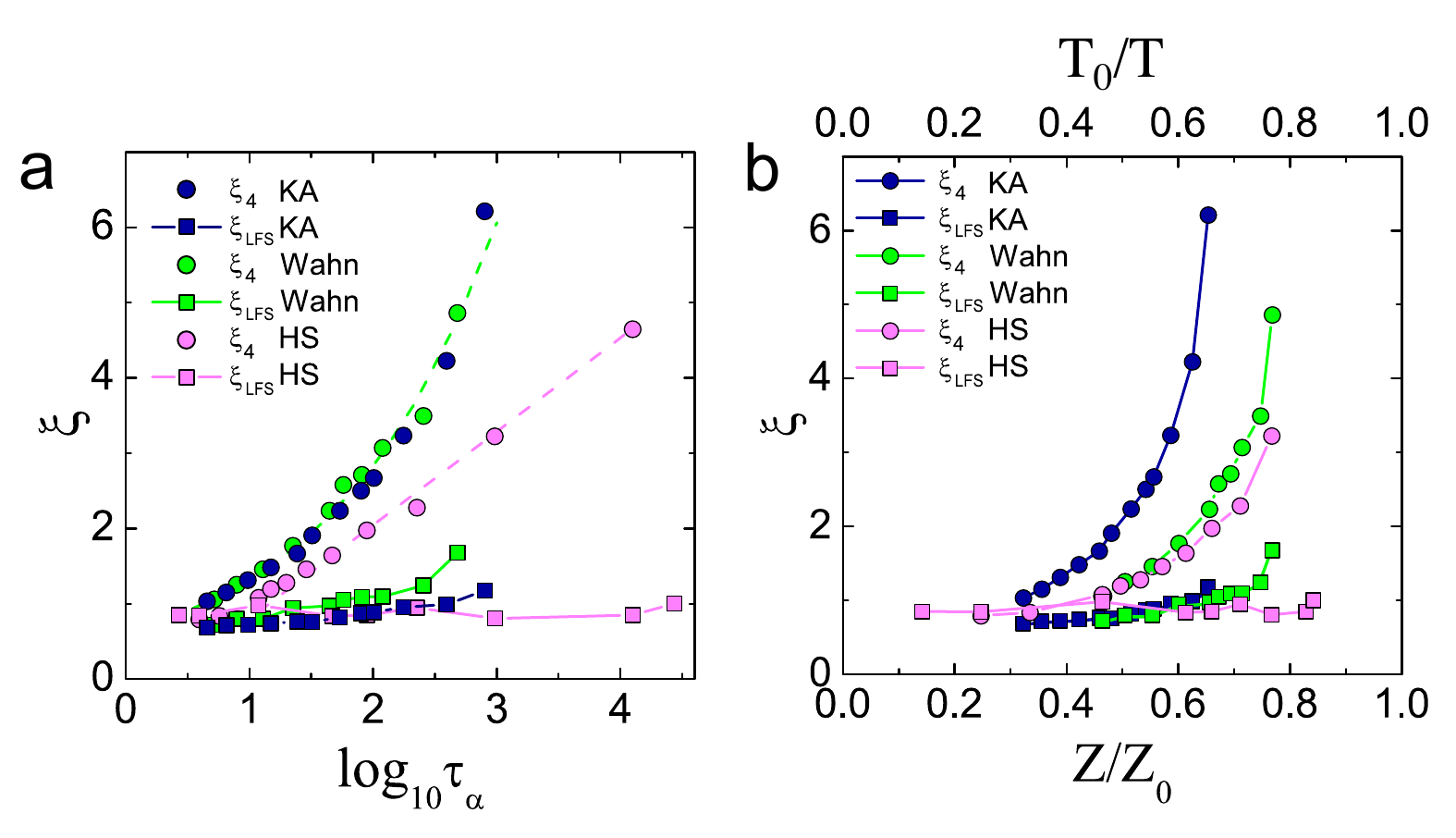}
 \caption{\textbf{Dynamic $\xi_4$ and static $\xi_\mathrm{LFS}$ lengths for the three systems considered.}
(a) $\xi_4$ and $\xi_\mathrm{LFS}$  as a function of $\tau_\alpha/\tau_0$.
(b) $\xi_4$ and $\xi_\mathrm{LFS}$ as a function of $T_0/T$ and $Z/Z_0$ for the Lennard-Jones and hard sphere systems respectively. In both figures, squares correspond to static length scales and circles to dynamic lengthscales. In (a) the dashed lines correspond to $\xi_4 \sim (\tau_\alpha^A)^{\gamma}$
with $\gamma=0.3$ and $\xi_4 \sim \ln(\tau_\alpha)$ for KA (green) and hard spheres (pink) as indicated.
In (a) $\tau_\alpha/\tau_0$ are offset for clarity.}
\label{figXiData}
\vspace*{-12pt}\end{center}
\end{figure*}

Example fits to Eq. \ref{eqOZ} are shown in Fig. \ref{figS}a for the Wahnstr\"{o}m model. The resulting $\xi_4$ are plotted in Fig. \ref{figXiData}. We see in Fig. \ref{figXiData}(a) that, as a function of $\tau_\alpha/\tau_0$, the $\xi_4$ for both Lennard-Jones systems coincide. The dynamic correlation length for the hard spheres rises more slowly across a wide range of $\tau_\alpha/\tau_0$. As was the case with the population of LFS [Fig. \ref{figNLFSN}(b)], plotting $\xi_4$ as a function of the degree of supercooling reflects the difference in fragility between these systems [Fig. \ref{figXiData}(b)]. For the Kob Andersen mixture  $\xi_4$ rises at comparatively weak supercooling, for hard spheres much more supercooling is needed to see a change in $\xi_4$.

Now the scaling of $\xi_4$ with relaxation time has been examined previously. In the case of the Wahnstr\"{o}m mixture, La\u{c}evi\'{c} \emph{et al.} ~\cite{lacevic2003} found behaviour consistent with divergence of $\xi_4$ at the temperature of the mode coupling transition. Whitelam  \emph{et al.} ~\cite{whitelam2004} obtained a $\xi_4$ scaling consistent with dynamic faclitation theory for the Kob-Andersen mixture. More recently, Flenner \emph{et al.}  ~\cite{flenner2009} found $\xi_4 \sim (\tau_\alpha)^{\gamma}$ with $\gamma \approx 0.22$ for the Kob-Andersen system. Kim and Saito have also found behaviour consistent with power law scaling for both  Kob-Andersen and Wahnstr\"{o}m mixtures ~\cite{kim2013}. In the case of hard spheres Flenner and Szamel found $\xi_4 \sim \ln(\tau_\alpha)$ ~\cite{flenner2011}. In Fig. \ref{figXiData} we find a slightly larger value for the exponent $\gamma\approx0.3$ for the Lennard-Jones models but find similar behaviour for hard spheres as that noted by Flenner \emph{et al.} ~\cite{flenner2011}. However, our hard sphere system is rather more monodisperse than the 1:1.4 binary mixture they used, which might account for the fact that our data is not entirely straight in the semi-log plot of Fig. \ref{figXiData}. Moreover hard spheres, and other systems do not always exhibit the same scaling for all $\tau_\alpha$ ~\cite{flenner2011,flenner2012}. In any case, we emphasise that such ``scaling'' is hard to assess on such small lengthscales (the entire range of $\xi_4$ is less than a decade), and extraction of reliable values for $\xi_4$ is far from trivial in finite-sized simulations ~\cite{flenner2007,karmakar2009}. We thus believe our finding of a differing exponent in the case of the Kob-Andersen mixture to that of Flenner and Szamel \emph{et al.} ~\cite{flenner2009} reflects the challenges of extracting such values. Furthermore, Flenner \emph{et al.} have recently extended their analysis to a wider range of systems which seem to indicate a common scaling of $\xi_4$ across a range of systems, including some identical or similar to those we consider which have a range of fragilities ~\cite{flenner2014}.

\textit{Static correlation length --- }
We now consider how to determine a static correlation length for the domains of locally favoured structures. It is clear from Fig. \ref{figPretty} that the LFS percolate. Given that all state points we have been able to access are necessarily far from $T_0$, and that the LFS themselves have a limited lifetime (Fig. \ref{figDTCC}), the existence of a percolating network of LFS does not itself imply arrest. However, as has been previously noted by others ~\cite{dzugutov2002,coslovich2007} and ourselves ~\cite{malins2013jcp,malins2013fara}, a percolating network of LFS has the potential to accelerate the increase of $\tau_\alpha$ upon supercooling. This is because the particles in the LFS act to slow down their neighbours and because domains of LFS last longer than isolated LFS ~\cite{malins2013jcp,malins2013fara}. However, identifying a lengthscale with the domain size of LFS, for example the radius of gyration, leads to divergence in the supercooled regime  ~\cite{malins2013jcp,malins2013fara}.

We thus turn to a method which allows a natural comparison with the dynamic lengthscale $\xi_4$. We define a structure factor restricted to the particles identified within LFS:
\begin{equation}
S_\mathrm{LFS}(\textbf{k})=\frac{1}{N\rho} \langle \sum_{j=1}^{N_\mathrm{LFS}} \sum_{l=1}^{N_\mathrm{LFS}} \exp[-i \textbf{k} \cdot \textbf{r}_j(t_0)]\exp[i \textbf{k} \cdot \textbf{r}_l(t_0)] \rangle,
\label{eqSLFS}
\end{equation} 
\noindent where $N_\mathrm{LFS}$ is the number of particles in LFS. We then fit the Ornstein-Zernike equation (Eq. \ref{eqOZ}) to the low-$k$ behaviour of the angularly-averaged $S_\mathrm{LFS}(k)$ in order to extract a structural correlation length $\xi_{\mathrm{LFS}}$.  This procedure is akin to the calculation of the dynamic lengthscale $\xi_4$: first a structure factor is calculated from a selected fraction of the particles (either immobile or structured), and the Ornstein-Zernike expression used to extract a correlation length  as shown in  Fig. \ref{figS}(b).

These $\xi_{\mathrm{LFS}}$ are plotted in Fig. \ref{figXiData} for the three systems we study. Like the dynamic correlation lengths,  the structural correlation lengths increase upon cooling for the Lennard-Jones systems, while the hard spheres show little change upon compression. However the manner in which these lengths increase is quite different. The main result is that the growth in the dynamic correlation length $\xi_4$ is not matched by the growth in the structural correlation length $\xi_{\mathrm{LFS}}$. Indeed $\xi_{\mathrm{LFS}} \sim \sigma$ through the accessible regime. This is less that the apparent thickness of the domains in Fig. ~\ref{figPretty}, however we note that $\xi_{\mathrm{LFS}}$ follows a reciprocal space analysis and that it may not directly correspond to the real space images. In any case the difference is only a factor of 2-3.

\section{Discussion}
\label{sectionDiscussion}

\textit{Geometric frustration is strong in model glassformers --- } Figure \ref{figXiData} provides a key result of this work. The correlation length related to the domains of locally favoured structures is short, around one particle diameter. Reference to Eq. \ref{eqCNTFrustration} and Fig. \ref{figFrustration} indicates that, according to our linear measure, geometric frustration is strong in these systems, in other words that $\xi_D^*=\xi_{\mathrm{LFS}}\approx\sigma$. 

Frustration has been demonstrated elegantly in curved space in 2d, where it has been controlled by the degree of curvature  ~\cite{sausset2008,sausset2010,sausset2010pre}.
However in 3d, the discussion involving 120 particles embedded on the surface of a four-dimensional sphere formed perfect icosahedra ~\cite{coexeter,coexeterpolytope} assumes these are monodisperse spheres. In addition to the fact that monodisperse spheres are usually poor glassformers, we have demonstrated here that very often, the LFS are not icosahedra. Moreover, even in the case of the Wahnstr\"{o}m mixture, it is far from clear that the icosahedra formed would tessellate the surface of a hypersphere with no strain as they are comprised of particles of differing sizes. 

We suggest that other curved space geometries may enable binary Lennard-Jones models to tessellate without strain. Alternatively, simulations in curved space of a one-component glassformer, such as Dzugutov's model ~\cite{dzugutov2002} may enable frustration to be investigated in 3d. Alternatively, controlling frustration with composition relates to work carried out by Tanaka and coworkers ~\cite{kawasaki2010jpcm,leocmach2012} which emphasises the role of medium-ranged crystalline order. However, unlike geometric frustration where the LFS are amorphous structures which form in the liquid and do not tessellate over large distances [Eq. \ref{eqCNTFrustration}], the medium-ranged crystalline order is distinct from the liquid, at least in $d=3$ ~\cite{tanaka2012}. A further comment to be made here concerns our identification of 10B clusters in the hard sphere system we consider, which is at odds with the local crystalline order found in hard spheres ~\cite{kawasaki2010jpcm,leocmach2012}. At present this discrepancy is being investigated. Pending the results of further analysis, we make the following observations. The TCC analysis has shown no indication of significant quantities of crystal like structures. It has been demonstrated that the TCC successfully detects FCC and HCP crystals and Lennard-Jones ~\cite{malins2013tcc} and hard spheres ~\cite{taffs2013}. We thus speculate that the bond-order parameter threshold used in ~\cite{kawasaki2010jpcm,leocmach2012} may allow some particles classified as 10B by the TCC to be interpreted as crystal-like order.

\textit{Fragility and structure --- } A considerable body of work suggests a link between fragility and the tendency of glassformers to develop local structure ~\cite{ito1999}. Strong and network liquids, such as silica, tend not to show large changes in local structure upon cooling ~\cite{coslovich2009}, although edge-sharing tetrahedra have been associated with fragility ~\cite{wilson2009}. In 2d, significant correlation between fragility and tendency to locally order is found ~\cite{shintani2006,kawasaki2007}. Recently, the development of multitime correlations has identified new timescales of dynamic heterogeniety. In fragile systems (in particular the Wahnstr\"{o}m mixture), this becomes much longer than $\tau_\alpha$ at deep supercoolings ~\cite{kim2013}.

In 3d similar behaviour is found in metallic glassformers ~\cite{mauro2012,mauro2013}, where in addition, glass-forming ability is associated with strong behaviour. However in model systems, such as hard spheres, in 3d at best only a very weak correlation between glass-forming ability (polydispersity) and fragility is found ~\cite{zaccarelli2009}. Moreover, systems with almost identical two-body structure can exhibit different fragilities ~\cite{berthier2009}, although higher-order structure in the form of LFS is correlated with fragility ~\cite{coslovich2011}. Conversely, we have shown that, in systems with effectively identical fragility, the change in structure upon cooling need not be same ~\cite{malins2013isomorph}. In higher dimension, structure becomes less important, but fragile behaviour persists ~\cite{charbonneau2013}. Finally, some kinetically constrained models, which are thermodynamically equivalent to ideal gases by construction, exhibit fragile behaviour ~\cite{pan2004}.

These observations make it clear that the development of local structural motifs upon supercooling is not always connected with fragility. These caveats aside, the data presented here in Fig. \ref{figNLFSN} for the three systems we have studied do suggest that more fragile systems show a stronger change in local structure upon supercooling. In particular, the Kob-Andersen model, which is the least fragile of the mixtures we consider, shows a continuous rise in bicapped square antiprisms across a wide range of temperatures. This is in marked contrast to the Wahnstr\"{o}m mixture mixture and hard spheres, both of which show a sharper rise in LFS population. We note that similar behaviour has been observed previously for the two Lennard-Jones mixtures  ~\cite{coslovich2007}.

\textit{Outlook --- }
Our work paints a picture of decoupling between structural and dynamic lengthscales in the simulation-accessible regime, which covers the first five decades of increase in structural relaxation time $\tau_\alpha$. By comparison, as shown in Fig. \ref{figAngell}, the molecular glass transition at $T_g$ corresponds to some 15 decades of increase in relaxation time. That the structural lengthscale decouples so strongly from the dynamic lengthscale $\xi_4$ suggests that larger regions than those associated with the LFS are dynamically coupled. As noted above, similar results have been obtained previously, using a variety of different measures
~\cite{hocky2012,charbonneau2012,kob2011non,dunleavy2012,charbonneau2013,malins2013jcp,malins2013fara}.

\begin{figure}[!htb]
\begin{center}
\includegraphics[width=55mm]{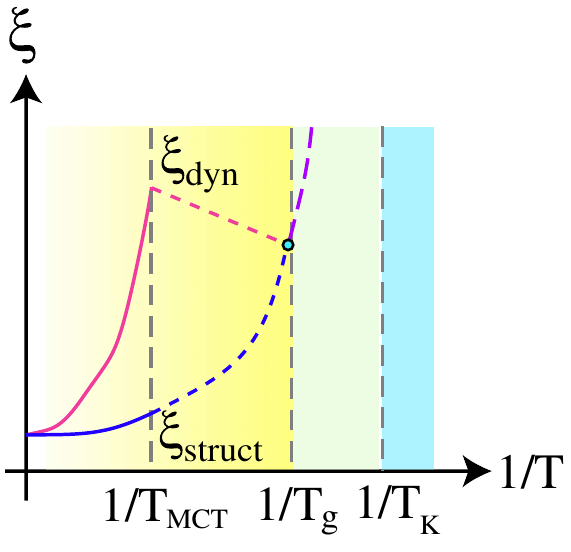} 
\caption{\textbf{Schematic of the possible behaviour of dynamic ($\xi_{\mathrm{dyn}}$) and static ($\xi_{\mathrm{stat}}$) lengthscales as the glass transition is approached.} Particle-resolved studies data from colloid experiment and computer simulation is available for $T\gtrsim T_{\mathrm{MCT}}$ (solid lines).
Dashed lines represent a possible scenario at lower temperatures, extrapolated from recent simulations ~\cite{kob2011non}. Green dot is dynamic length $\xi_3$ deduced from molecular experiments close to $T_g$ \cite{berthier2005}. Purple dashed line represents coincidence of structural and dynamic lengths of cooperatively re-arranging regions envisioned by RFOT and Adam-Gibbs theory.}
\label{figXi}
\vspace*{-12pt}\end{center} 
\end{figure}

The picture that emerges is one of disparity between $\xi_4$ and the majority of structural lengths, as illustrated in Fig. \ref{figXi}. This leaves at least three possibilities:

\begin{enumerate}
\item Dynamic and structural lengths decouple as the glass transition is approached. And thus although structural changes are observed in many fragile glassformers, they are not a mechanism of arrest.
\item $\xi_4$ is not representative of dynamic lengthscales, or its increase as a function of supercooling is not sustained.
\item The majority of data so far considered is in the range $T>T_\mathrm{MCT}$ and thus is not supercooled enough for RFOT or Adam-Gibbs-type cooperatively re-arranging regions dominate.
\end{enumerate}

We believe that a combination of all three, weighted differently depending on the model, is the most likely outcome. Some evidence for the first scenario is given by the fact the kinetically constrained models ~\cite{pan2004}, and hyperspheres in high dimension ~\cite{charbonneau2013} undergo arrest. If one accepts either of these (admittedly abstract) models, structure cannot be a universal mechanism for dynamical arrest. 

Further evidence in support of scenario one is provided by Cammarota and Biroli ~\cite{cammarota2012} that pinning a subset of particles can drive the ideal glass transition of the type envisioned by random first order transition theory. Under the pinning field, no change in structure occurs (subject to certain constraints) as a function of pinned particles, but a bona-fide glass transition as described by RFOT theory does  ~\cite{cammarota2012}. One possibility is to note that, as temperature drops, a lower concentration of pinned particles is required for this pinning glass transition and that the transition is somehow driven by a combination of structure and pinning. Moreover, the separations between the pinned particles in the simulation accessible regime can approach one or two particle diameters ~\cite{kob2013}, suggesting rather small cooperatively re-arranging regions.

However the comparatively rapid increase in $\xi_4$ is not without question. Firstly, as indicated in Fig. \ref{figXi}, $\xi_4$ seems to increase rapidly in the regime accessible to particle-resolved studies in the regime $T \gtrsim T_{\mathrm{MCT}}$ ~\cite{harrowell2011}. Indeed, a free fit to measurements of $\xi_4$ for our data using the Kob-Andersen model yielded \emph{divergence} close to the Mode-Coupling temperature ~\cite{malins2013fara}, which may be an ``echo'' in $d=3$ of the mean-field MC transition. We find the ``critical exponent'' is $\nu =0.588 \pm 0.02$, the ``critical temperature'' is $T_\mathrm{C}=0.471 \pm 0.002$ and the prefactor is $\xi_4^0 =0.59 \pm 0.02$. Under the caveat that obtaining $\xi_4$ from fitting $S_4$ in limited size simulations is notoriously problematic~\cite{flenner2007,flenner2009} and thus any numerical values should be treated with caution, we observe that the value of $T_\mathrm{C}$ is not hugely different to the  transition temperature found by fitting Mode-Coupling theory to this system, around 0.435~\cite{kob1995a,kob1995b}. We also note that $\nu =0.588$ lies between mean field ($\nu =0.5$) and 3D Ising ($\nu =0.63$) criticality. Moreover among early papers involving $\xi_4$, La\v{c}evi\'{c} \emph{et al.} showed divergence of this dynamic length around $T_\mathrm{MCT}$. Furthermore, a recent paper by Kob \emph{et al.} ~\cite{kob2011non} indicates \emph{non-monotonic} behaviour of a dynamic correlation length based on pinning with a maximum around $T_{\mathrm{MCT}}$ as indicated in Fig. \ref{figXi}. These results are not without controversy ~\cite{flenner2012comment,kob2012reply}, but it has since been shown that, just below $T_\mathrm{MCT}$, at the limit of the regime accessible to simulations, $\xi_4$ can tend to saturate ~\cite{mizuno2011} and at least exhibits a different scaling ~\cite{flenner2013}. 

Additional evidence that the dynamic correlation might not diverge as fast as data from the $T> T_{\mathrm{MCT}}$ range might indicate is given by experiments on molecular glassformers close to $T_g$, some 8-10 decades increase in relaxation time compared to the particle-resolved studies. This approach measures a lower bond for the dynamic correlation length ~\cite{berthier2005}. The lengths obtained by this approach correspond to a few molecular diameters ~\cite{berthier2005,crauste2010,brun2011}. Such a small dynamic correlation length (albeit a lower bound) certainly necessitates at the very least a slowdown in the rate of increase of $\xi_{\mathrm{dyn}}$ followed by a levelling off. 
Finally, we emphasise that $\xi_4$ may not be the only means to define a dynamic length ~\cite{harrowell2011,kob2011non}.

Another possibility is to question the decomposition of the complex geometries indicated in Fig. \ref{figPretty} onto a single linear measure. Clearly this is a simplification and one which is not necessarily justified. Indeed these networks percolate, which implies that the radius of gyration of the domains of LFS must diverge. That percolation of LFS occurs at such moderate supercooling indicates that divergence of the radius of gyration of domains does not lead to arrest \cite{malins2013jcp}. However other analyses of the LFS network may provide further insight.

It is tempting to imagine that in the $T_{\mathrm{MCT}}>T>T_g$ range (or even in the regime below $T_g$), structural and dynamic lengths might scale together, corresponding to well-defined cooperatively re-arranging regions (Fig. ~\ref{figXi}). The discussion of geometric frustration, and in particular Eq. \ref{eqCNTFrustration} suggest that an increase in structural lengthscale might necessitate either a decrease in frustration, or ``surface tension'' or the thermodynamic driving force to form locally favoured structures. Calculating any of these quantities, given the short lengthscales and complex geometries involved appears a formidable task, but which might provide a framework for increasing structural lengthscales at deep supercooling.

For now, however, the jury is well and truly out as to the nature of any structural mechanism for dynamic arrest. Locally favoured structures can be identified and form networks which might at deeper supercooling ($T<T_{\mathrm{MCT}}$) lead to the emergence of solidity in glassforming liquids. Hints in this direction are evidenced from the growth in LFS with supercooling, that particles in LFS are slower than average and that they retard the motion of neighbouring particles ~\cite{malins2013jcp,malins2013fara} although the degree to which LFS \emph{predict} the dynamics in the accessible regime is limited ~\cite{jack2014}. However the discrepancy observed by some between structural and dynamic lengthscales in the $T \gtrsim T_{\mathrm{MCT}}$ range is indicative that more is at play than structure at least in the first few decades of dynamic slowing which are described by mode-coupling theory.

\section*{Conclusions} 
\label{sectionConclusions}

In three model glassformers, we have identified the locally favoured structure with the dynamic topological cluster classification. Each system exhibits a distinct LFS, which lasts longer than all other structures considered : the 11A bicapped square antiprism in the case of the Kob-Andersen model, the 13A icosahedron for the Wahnstr\"{o}m mixture and the 10B cluster for hard spheres. In all these systems, the LFS form a percolating network upon supercooling in the simulation accessible regime. In this dynamical regime accessible to simulation, the formation of this network does not correlate with dynamic arrest : all our systems continue to relax after a percolating network of LFS has formed. The network formation is qualitatively similar in all systems, although the less fragile Kob-Andersen mixture exhibits a less dramatic rise in LFS population than either the Wahnstr\"{o}m mixture or hard spheres. We investigate structural and dynamic lengthscales. In all cases the dynamic length $\xi_4$ increases much faster than the structural length in the dynamic regime accessible to our simulations. The lack of growth of the structural correlation length appears compatible with strong geometric frustration.

\section{Acknowledgements} 
The authors would like to thank C. Austen Angell, Giulio Biroli, Patrick Charbonneau, Daniele Coslovich, Jens Eggers, Peter Harrowell, Rob Jack, Ken Kelton, Matthieu Leocmach, Tannie Liverpool, Grzegorz Szamel, Hajime Tanaka, Gilles Tarjus, Stephen Williams and Karoline Weisner for many helpful discussions. CPR would like to thank Daniel Crespo and Kia Ngai for their kind help during his Barcelona excitements. CPR would like to acknowledge the Royal Society and the ERC ``Nano-PRS'' for financial support. A.M., A.J.D. and R.P. are funded by EPSRC grant code EP/E501214/1. This work was carried out using the computational facilities of the Advanced Computing Research Centre, University of Bristol.


%

\end{document}